\documentclass[]{spie}  
\usepackage[]{times} 
\usepackage[]{graphicx}
\usepackage{textcomp}
\usepackage{hyperref}

\title{PRAXIS - a low background NIR spectrograph for fibre Bragg grating OH suppression} 

\author{Anthony Horton\supit{a}, Simon Ellis\supit{a}, Jon
  Lawrence\supit{a} and Joss Bland-Hawthorn\supit{b} 
\skiplinehalf
\supit{a}Australian Astronomical Observatory, PO Box 296, Epping NSW
1710, Australia; \\ 
\supit{b}Institute for Astronomy, School of Physics, University of
Sydney NSW 2006, Australia 
}

\authorinfo{Send correspondence to ajh@aao.gov.au}

\begin{document} 
  
Anthony Horton et al, "PRAXIS: a low background NIR spectrograph for fibre Bragg grating OH suppression", {\em Modern Technologies in Space- and Ground-based Telescopes and Instrumentation II}, {\em Proc.~SPIE}~{\bf 8450}, 84501V (2012)

Copyright 2012 Society of Photo Optical Instrumentation Engineers. One print or electronic copy may be made for personal use only. Systematic electronic or print reproduction and distribution, duplication of any material in this paper for a fee or for commercial purposes, or modification of the content of the paper are prohibited.

\url{http://dx.doi.org/10.1117/12.924874}

  \pagebreak

  \maketitle 

\begin{abstract}
Fibre Bragg grating (FBG) OH suppression is capable of greatly
reducing the bright sky background seen by near infrared
spectrographs. By filtering out the airglow emission lines at high
resolution before the light enters the spectrograph  this technique
prevents scattering from the emission lines into interline regions,
thereby reducing the background at all wavelengths. In order to take
full advantage of this sky background reduction the spectrograph must
have very low instrumental backgrounds so that it remains sky noise
limited. Both simulations and real world experience with the prototype
GNOSIS system show that existing spectrographs, designed for higher
sky background levels, will be unable to fully exploit the sky
background reduction. We therefore propose PRAXIS, a spectrograph
optimised specifically for this purpose. 

The PRAXIS concept is a fibre fed, fully cryogenic, fixed format
spectrograph for the J and H-bands. Dark current will be minimised by
using the best of the latest generation of NIR detectors while thermal
backgrounds will be reduced by  the use of a cryogenic fibre
slit. Optimised spectral formats and the use of high throughput volume
phase holographic gratings will further enhance sensitivity. Our
proposal is for a modular system, incorporating exchangeable
fore-optics units, integral field units and OH suppression units, to
allow PRAXIS to operate as a visitor instrument on any large telescope
and enable new developments in FBG OH suppression to be incorporated
as they become available. As a high performance fibre fed spectrograph
PRAXIS could also serve as a testbed for other astrophotonic
technologies.
\end{abstract}



\section{INTRODUCTION}
\label{sec:intro}  

Fibre Bragg grating OH suppression is a high performance optical
filter technology.  A fibre Bragg grating (FBG) is a length of single
mode fibre with refractive index variations along the axis which cause
specific wavelengths to be reflected.  Aperiodic FBGs can be used to
create multi-notch filters with large numbers of narrow, deep, square
profiled  notches while retaining high interline throughput, exactly
the properties required for effective filtering of the OH emission
lines that dominate the near infrared sky
background\cite{Bland-Hawthorn2004}.  The current state of the art
allows for over 100 notches at a  resolution of $R\approx 10000$,
depths of over 30dB, profiles well fit by $n=5$ Butterworth functions
and interline throughput of over 90\% \cite{Trinh2012}.  The single
mode fibres (SMFs) required for FBGs generally cannot be used directly
in astronomical applications because of low coupling efficiencies
however devices known as photonic lanterns (PLs) enable arrays of FBGs
(or other single mode photonic devices) to be used be used with
multimode input and output\cite{Leon-Saval2005a}.

FBG OH suppression is capable of greatly reducing the bright sky
background observed by near infrared spectrographs. By filtering out
the OH emission lines before the light enters the spectrograph this
technique prevents the scattering of light from the emission lines
into interline regions and as a result has the ability to reduce the 
background at all wavelengths\cite{Ellis2008}. 
The effectiveness of this background suppression clearly has the
potential to significantly increase the sensitivity of ground based
near infrared spectroscopy however in order to fulfil this potential
the spectrographs must have very low instrumental backgrounds so that
they remain sky background noise limited.  Simulations as well as real
world experience gained during the commissioning of the GNOSIS system
with the IRIS2 spectrograph on the AAT\cite{Ellis2012,Trinh2012}
suggest that existing spectrographs, optimised for much higher sky
background levels, will be unable to fully exploit the benefits FBG OH
suppression.  A purpose built, specialised spectrograph is required 
and for that reason we plan to built PRAXIS.

\section{DESIGN CONSIDERATIONS} 

The design of a spectrograph intended for use with FBG OH
suppression must consider several key factors.


   \begin{figure}
   \begin{center}
   \includegraphics[width=0.9\textwidth]{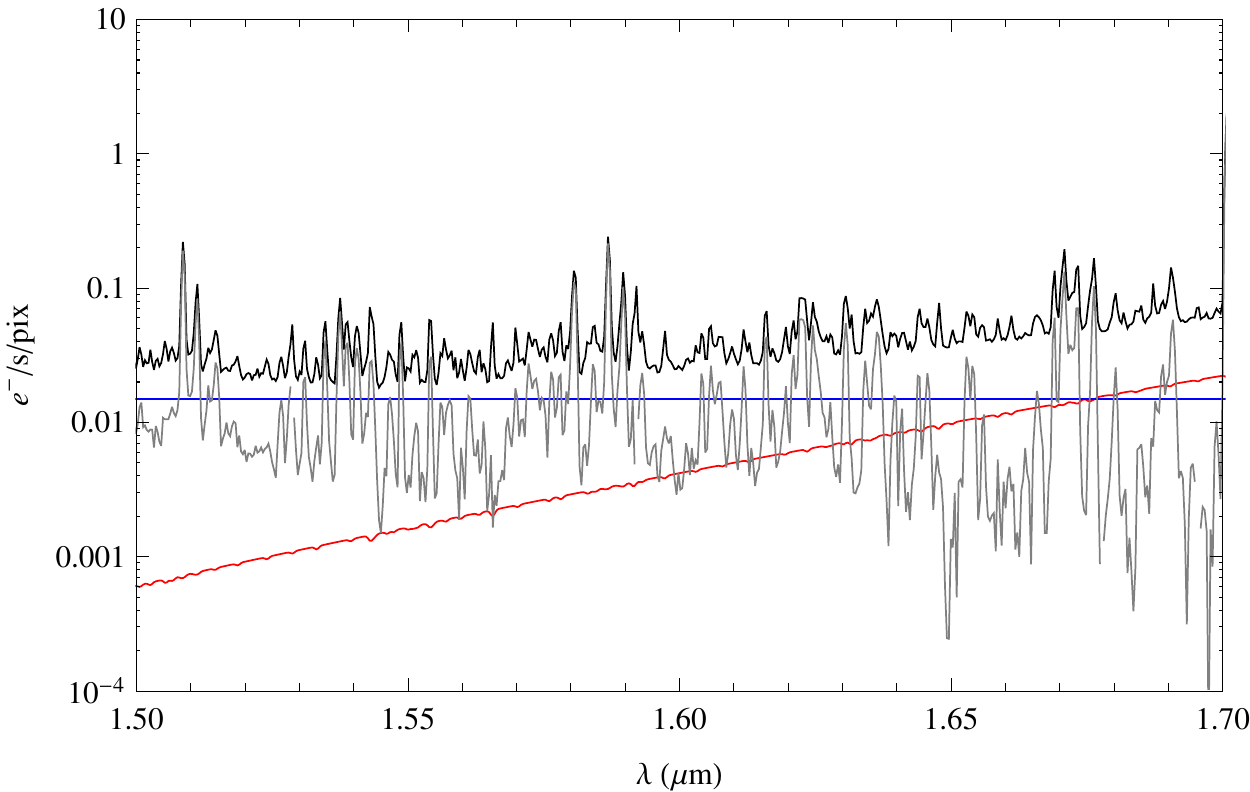}
   \end{center}
   \caption[example] 
   { \label{fig:GNOSIS_background} 
Observed background of the GNOSIS prototype FBG OH suppression system, taken
from Ellis et al 2012\cite{Ellis2012}.  The black line shows the total
background, the horizontal blue line is the detector dark current, the
red line is the instrument thermal background and the gray line shows
the residual background after substraction of the dark current and
instrument thermal components.}
   \end{figure}

\subsection{Dark Current} 
\label{sec:title}

Simulations comparing the various sources of background signals in FBG
OH suppressed spectrographs predict that detector dark 
current will be a significant contributor and care must be taken to minimise it in order to prevent the associated Poisson 
noise limiting the sensitivity of the instrument.  Results from
GNOSIS, a prototype FBG OH suppression system which used the existing
IRIS2 spectrograph on the Anglo-Australian Telescope, illustrate the potential impact of dark current.  Figure \ref{fig:GNOSIS_background} shows the background levels observed by GNOSIS, as reported by Ellis et~al\cite{Ellis2012}.  The PACE HAWAII-1 HgCdTe detector used in IRIS2 has a measured dark current of 0.015 e$^-$s$^{-1}$pixel$^{-1}$ however even this relatively low value is sufficient to dominate the observed background for wavelengths less than 1.67 \textmu m.   There are two main factors effecting the impact of dark current, detector selection and spectral format.

There are large differences in dark current levels between different near infrared detector arrays, InSb and first generation (PACE) HgCdTe detectors have dark currents ranging from a few e$^-$s$^{-1}$pixel$^{-1}$ to $\sim 10^{-2}$ e$^-$s$^{-1}$pixel$^{-1}$. The current generation of MBE HgCdTe detectors are 
able to achieve significantly lower dark currents, however.  Dark currents of $\sim 2\times 10^{-3}$ e$^-$s$^{-1}$pixel$^{-1}$ have been measured for 2.5 \textmu m cutoff devices and it is expected that with sufficient cooling dark currents of $\sim 10^{-3}$ e$^-$s$^{-1}$pixel$^{-1}$ are possible for the 1.75 \textmu m cutoff detectors\cite{Blank2011}.  The majority of infrared astronomical spectrographs currently in service use InSb or PACE HgCdTe detectors but without the order of magnitude reduction in dark current offered by a MBE HgCdTe detector it would be difficult for a spectrograph with FBG OH suppression to avoid being detector noise limited.

The spectral format on the detector is also important, minimising the number of pixels over which the signal from the 
sky is spread minimises the contribution of dark current.  There are three aspects to this, all of which should be followed by an optimised spectrograph.  First, the spectral resolution should be the minimum that meets the science requirements for the instrument.  Note that because the OH lines are filtered out at high resolution ($R\sim 10000$) before entering the spectrograph an instrument with FBG OH suppression is able to use lower resolutions than a spectrograph without OH suppression provided that the resolution is sufficient to achieve the science objectives.  Second, the pixel scale of the spectrograph should be chosen to just critically sample the spectral resolution element, i.e.\ the fibre images should subtend a diameter of $\sim2.5$ pixels on the detector.  Third, the spectrograph should minimise the number of pixels in the spatial direction.

There is another factor effecting the impact of detector noise on the spectrograph and that is the A$\Omega$ product or \'{e}tendue per fibre, the higher the A$\Omega$ the higher the surface brightness on the detector and the higher the ratio between signal and dark current.  The A$\Omega$ per fibre is not usually determined by the spectrograph design however, it is generally constrained by other factors.  At the input of the observing system the A$\Omega$ per fibre can be equated to the product of the collecting area of the telescope and the solid angle on the sky that each fibre aperture projects to.  The former is clearly fixed for a given telescope while the latter will often be determined by the typical angular size of the targets that will be observed, or the features of interest within them.  In the case of the FBG OH suppression there is also a technological constraint on the A$\Omega$ per fibre due to the properties of photonic lanterns.  A photonic lantern will only transmit light below a maximum A$\Omega$ which is determined by the number of SMFs in the lantern.  For a given photonic lantern the maximum field of view per fibre on sky, $\theta_{max}$, can be approximated by
\begin{equation}
\theta_{max} \sim 4 \sqrt{N} \lambda / \pi D_{tel}
\end{equation}
where $N$ is the number of SMFs, $\lambda$ is the wavelength and $D_{tel}$ is the telescope diameter.  The GNOSIS FBG OH suppression system has 19 SMFs per photonic lantern, operates over the wavelength range 1.47--1.70 \textmu m and was installed on the 3.9 m AAT, this gives a maximum field of view per fibre
of approximately 0.4 arcseconds.  This is somewhat smaller than ideal given the typical seeing at the site however budget and timescale constraints limited the size of the photonic lanterns that could be considered at the time.  Photonic lanterns with 61 SMFs, which would allow $1.8\times$ the A$\Omega$ per fibre, have already been demonstrated\cite{Noordegraaf2010a} and developments in multicore fibre and integrated optics photonic lanterns promise to make large values of $N$ readily achievable\cite{Min2012,Haynes2012,Spaleniak2012} in the near future.  A spectrograph designed for FBG OH suppression should have collimator and camera optics with sufficiently fast focal ratios to accommodate the beam angles associated with the maximum A$\Omega$ values of the photonic lanterns it will be used with.

\subsection{Thermal Background} 

As with dark current low levels of instrument thermal background are required in order for a spectrograph to remain sky noise limited with FBG OH suppression.  As can been seen in figure \ref{fig:GNOSIS_background} the GNOSIS FBG OH suppression system exhibited a significant instrument thermal background component, high enough to be the dominant source of background for wavelengths longer than 1.67 \textmu m.  The main source of this background is believed to be H-band thermal emission from the fibre slit.  Because GNOSIS was fitted to an existing spectrograph that was not designed to be fibre fed it was necessary to use a fibre slit outside the spectrograph dewar and feed the spectrograph via an optical relay.  The thermal emission from the uncooled fibre slit is difficult to effectively baffle without vignetting the light path from the fibres themselves.  An infrared spectrograph designed to be fibre fed from the outset can substantially reduce instrument thermal background by positioning the fibre slit inside the dewar.  This eliminates the need for a dewar window, reduces the number of warm optical elements in the optical path, and of course greatly reduces emission from the fibre slit itself.

\subsection{Throughput} 

High throughput is obviously advantageous for any spectrograph but this is especially true in the case of intrinsically low signal levels 
combined with significant levels of instrument background (dark current and thermal).  The most important factor in 
determining the throughput of the spectrograph is the choice of dispersive element, for example the surface relief grism used in IRIS2 has an H-band peak diffraction efficiency of approximately 40\%\ whereas volume phase holographic gratings would be expected to give over twice that.  Compared to general purpose near infrared spectrographs an instrument designed for FBG OH suppression can make further throughput gains by optimising the optical design and coatings for a single format and reduced wavelength range (J \&\ H-bands only).

\subsection{Versatility} 

Fibre Bragg grating OH suppression systems are still under active development, therefore important a spectrograph 
that can be used to test a number of different OH suppression units is needed. Likewise in order to maximise the opportunities for telescope 
time the spectrograph should be usable with any large optical telescope. These requirements suggest a modular approach with 
interchangeable front end components (fore optics, integral field units, OH suppression units) which allow the spectrograph 
itself to be used unmodified on different telescopes and with different OH suppression unit designs. The spectrograph itself 
should be largely self-sufficient in operation, requiring only standard services such as electrical power and coolant supplies.

\section{INSTRUMENT CONCEPT} 

   \begin{figure}
   \begin{center}
   \includegraphics[width=0.8\textwidth]{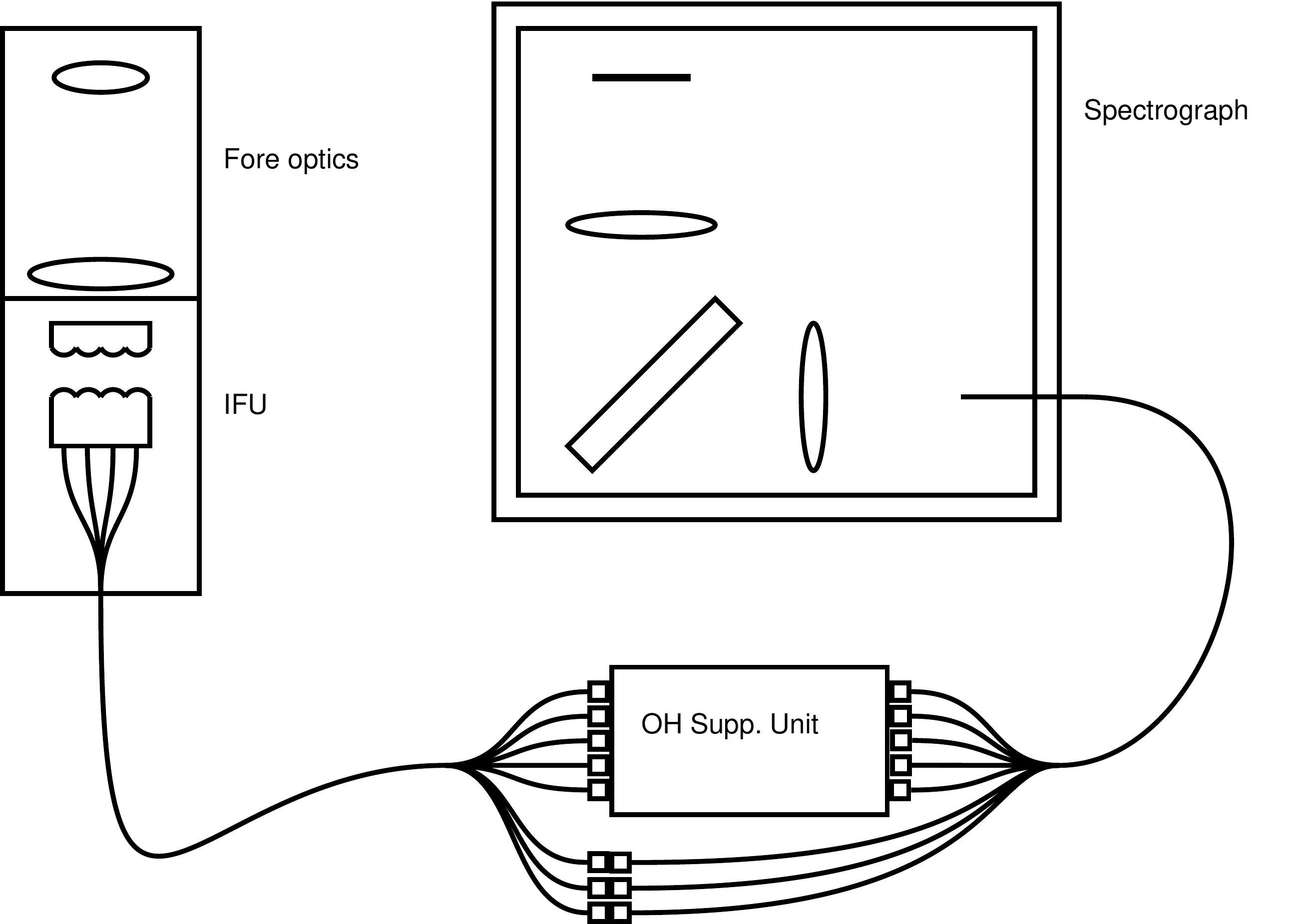}
   \end{center}
   \caption[example] 
   { \label{fig:PRAXIS_schematic} 
Schematic showing the components of the PRAXIS modular instrument system.}
   \end{figure}

\begin{figure}
   \begin{center}
   \includegraphics[width=\textwidth]{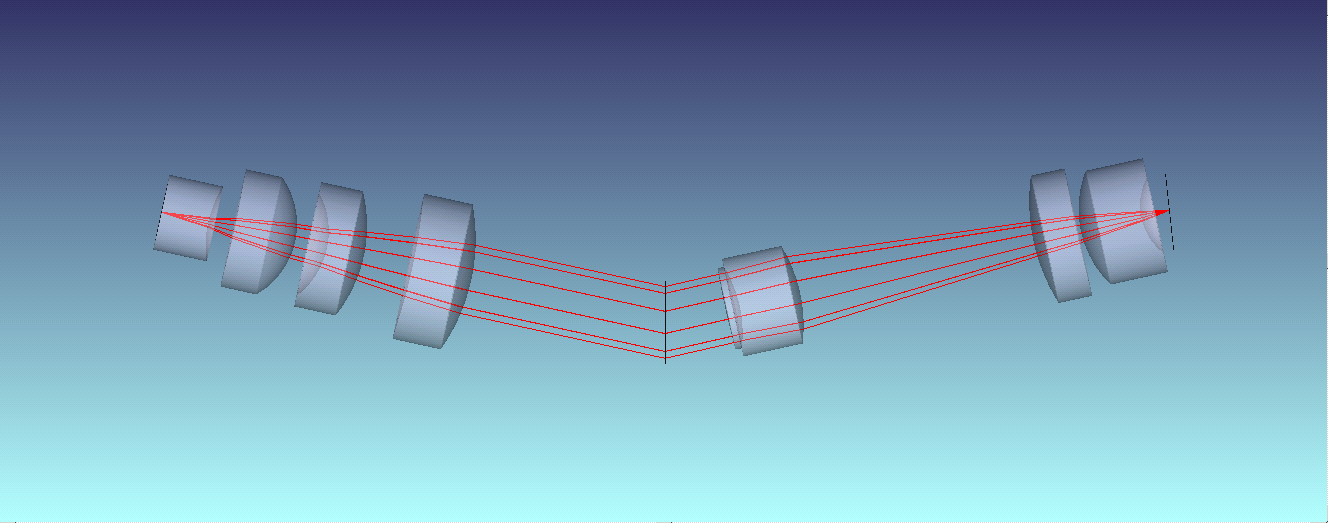}
   \end{center}
   \caption[example] 
   { \label{fig:PRAXIS_layout} 
 Optical design concept for the PRAXIS spectrograph.}
   \end{figure}

The PRAXIS instrument concept is illustrated in figure \ref{fig:PRAXIS_schematic}.  It consists of the following components:

\begin{description}
\item[Spectrograph.]{Design based around 50 \textmu m core diameter fibres with output focal ratios of up to f/3.  This gives a maximum field of view per fibre of 0.87''/0.43'' on 3.9/8 m telescopes and is well matched to the A$\Omega$ of photonic lanterns with up to $\sim 70$ SMFs at 1.55 \textmu m, or up to $\sim130$ at 1.15 \textmu m.  Larger field of view/SMF counts could be accomodated using larger diameter core fibres but at the expense of spectral resolution.

The spectrograph is a bench mounted, fixed format, fully cryogenic volume phase holographic grating based design.  The spectral resolution will be in the range $R=$1000-2000 (to be confirmed) and the operating wavelength range will be 1.45--1.70 \textmu m in the baseline version with a J-band upgrade path.  The detector will be a single MBE HgCdTe array with at least $1024\times1024$ pixels.  Both the collimator and camera will be f/3 designs, giving a fibre image diameter of 2.8 pixels (assuming 18 \textmu m pixels).  A transmissive optical design will be used to mimise the size (and cost) of the instrument.  Figure \ref{fig:PRAXIS_layout} illustrates an optical design concept for the spectrograph.  The required spectral resolution can be acheived with a beam size under 50 mm, allowing the collimator and camera focal lengths to be less than 150 mm and the overall sizto be only $\sim 300$ mm.

A gravity invariant location allows a simple optical table based mechanical design and makes LN$_2$ based cooling straightforward to implement. No cryogenic mechanisms are required and the only external connections are the vaccuum feedthrough of the fibre bundle and the electrical connections for the detector.
}

\item[Fibre bundle 1.]{Optical fibre bundle which connects the OH suppression unit to the spectrograph.  As the OH suppression unit and spectrograph will be colocated then this bundle only needs to be 1-2 m in length. The fibres will be connectorised at the OH suppression unit end. At the spectrograph end the fibres will pass through a vacuum feed through panel into the spectrograph dewar and terminate in a cryogenic fibre slit.  The fibres will be 50/125/150 \textmu m core/clad/buffer diameter with a polyimide buffer for vacuum compatibility.  The fibre
slits will be V-groove arrays with $\sim 180$\textmu m pitch allowing up to 100 fibres per 1k pixels.}

\item[OH suppression unit.]{An enclosure containing the photonic lanterns and fibre Bragg gratings which perform the OH suppression, to be co-located with the spectrograph at a gravity invariant location near to the telescope focal station. Fibre connectors will be required to connect fibre bundle 1 to the OH suppression unit outputs and to connect the OH suppression unit inputs to fibre bundle 2.}

\item[Fibre bundle 2.]{Optical fibre bundle which connects the OH suppression unit to the focal station.  The fibres are permanently bonded to an integral field unit (IFU) at one end and connectorised at the other to facilitate use of the fibre bundle with different OH suppression units. The fibre bundle is protected with an armoured limited bend flexible conduit.  The length should be sufficient to reach from likely telescope focal stations to the nearest gravity invariant locations, e.g. from the AAT Cassegrain focus to the dome floor beneath the horseshoe, from the Gemini ISS to the azimuth platform or from the VLT Nasmyth adaptor rotator to the Nasymth platform. Approximately 10--15 m should be sufficient for most telescopes.}

\item[Integral field unit.]{Microlens array which projects telescope pupil images onto the end faces of an optical fibre array.  This will use the 250 \textmu m pitch double microlens array telecentric design also planned for the KOALA\cite{EllisS.C.2012} and GHOST\cite{Ireland2012} instruments.  The image space focal ratio of the array should be matched to the acceptance cone of the photonic lanterns in the OH suppression units to be used with the IFU, therefore different IFUs would be required for optimal operation with different number of SMF lanterns.  The use of a demountable kinematic mount to connect the IFU to the fore optics unit would allow IFUs to be exchanged as required. The number of elements in the array should be sufficient to provide an on-sky field of view of several arcseconds in diameter so that the IFU can be used to reconstruct images suitable for target acquisition.}

\item[Fore-optics.]{Magnifying and field flattening optics mounted at the telescope focal station to convert the beam from the telescope to the focal ratio required by the integral field unit, which is determined by the ratio of the fibre core diameter to IFU effective focal length. If this ratio is fixed then the required magnification depends only on the focal ratio of the telescope. In that case two different sets of optics would be required to handle both the f/8 Cassegrain focus of the AAT and the approximately f/15 focal stations of 8-10 m telescopes.  The fore optics will also incorporate field and pupil stops to suppress scattered light.}

\end{description}

\section{STATUS \&\ PLANS}

At the time of writing the instrument concept is currently under development.  Funding is available for the spectrograph and fibre components however, and prototyping and testing of critical fibre components such as the vaccuum feedthroughs, cryogenic slit and fibre connectors will begin this year.  Several options are currently being investigated for obtaining access to a suitable HgCdTe detector.  The aim is for PRAXIS to be completed and commissioning in 2013.



\end{document}